\documentclass[twocolumn,showpacs,prl]{revtex4}

\usepackage{graphicx}%
\usepackage{dcolumn}
\usepackage{amsmath}
\usepackage{latexsym}

\begin {document}

\title
{
Scale-free network on a vertical plane
}
\author
{
S. S. Manna$^{1,2}$, G. Mukherjee$^{1,3}$ and Parongama Sen$^4$
}
\affiliation
{
$^1$Satyendra Nath Bose National Centre for Basic Sciences
    Block-JD, Sector-III, Salt Lake, Kolkata-700098, India \\
$^2$John-von-Neumann Institute for Computing, Forschungszentrum J\"ulich, D-52425 J\"
ulich, Germany \\
$^3$Bidhan Chandra College, Asansol 713304, Dt. Burdwan, West Bengal, India \\
$^4$Department of Physics, University of Calcutta, 92 Acharya Prafulla Chandra Road,
Kolkata 700009, India
}
\begin{abstract}

      A scale-free network is grown in the Euclidean space with a global 
   directional bias. On a vertical plane, nodes are introduced at unit 
   rate at randomly selected points and a node is allowed to be connected 
   only to the subset of nodes which are below it using the attachment 
   probability: $\pi_i(t) \sim k_i(t)\ell^{\alpha}$. Our numerical
   results indicate that the directed scale-free network for 
   $\alpha=0$ belongs to a different universality class compared to the 
   isotropic scale-free network. For $\alpha < \alpha_c$ the degree distribution
   is stretched exponential in general which takes a pure exponential
   form in the limit of $\alpha \to -\infty$. The link length distribution
   is calculated analytically for all values of $\alpha$.

\end{abstract}

\maketitle

       It has been seen in many branches of Statistical Physics that a
    global directional bias in space has strong effect on the critical 
    behaviour of simple models. Introduction of a preferred direction
    in the system reduces the degrees of freedom of the constituting elements
    of the system, which shrinks the configuration space available to the system
    compared to the undirected system. As a result a directed system is simpler
    and quite often tractable analytically. Examples include
    Directed percolation \cite {DP}, 
    Directed Sandpile Model \cite {DSM}, Directed River networks \cite {DRN}
    and Directed Self-avoiding walks \cite {DSAW} etc.
    
       Over the last few years it is becoming increasingly evident that highly
    complex structures of many social \cite {Newman}, biological \cite {Jeong,Sole}
    or electronic communication \cite {web,Faloutsos}
    networks etc. cannot be modeled by simple random graphs. For example
    in the well known random graphs by Erd\"os and R\'enyi, the degree
    distribution $P(k)$ is Poissonian (degree $k$ of a vertex is the number of edges
    attached to it) \cite {Erdos}.
    In contrast, it has been observed recently that the nodal degree distributions of 
    many networks, e.g., World-wide web \cite {web} 
    and the Internet \cite {Faloutsos} have power law tails: 
    $P(k) \sim k^{-\gamma}$. Due to the absence 
    of a characteristic value for the degrees these networks are 
    called `scale-free networks' (SFN) \cite {barabasi,linked,review,ves}.
    Barab\'asi and Albert (BA) generated scale-free graphs where
    a fixed number of vertices are added at each time and are 
    linked with a linear attachment probability \cite {barabasi}.
    On the other hand some of these networks are directed networks whose
    links are meaningful only when there is a connection from one
    end to the other but not the opposite, e.g., the
    World-wide web \cite {web}, the phone-call graph \cite {phone}
    and the citation graph \cite {Redner}.

      However, there are networks in which the nodes are geographically 
   located in different positions on a two-dimensional Euclidean space
   e.g., electrical networks, Internet or even in postal and transport
   networks etc. The edges of 
   the graphs representing these networks carry non-uniform weights which
   in most cases are either equal or proportional to the Euclidean lengths of the
   links. In these networks a relevant question is how to optimize
   the total cost of the connections e.g., electrical wires, Ethernet cables or say 
   travel distances of postal carriers \cite {Manna-Alkan}. On the other hand 
   a detailed knowledge
   of link length distribution is also important in the study of Internet's topological structure 
   for designing efficient routing protocols and modeling Internet 
   traffic. For example, Waxman model describes the Internet with exponentially 
   decaying link length distribution \cite {Waxman}. 
   Yook et. al. observed that nodes of the router level network maps of North America
   are distributed on a fractal set and the link length distribution is inversely 
   proportional to the link lengths \cite {yook1}. Other models of networks
   on Euclidean space are also studied in the literature 
   \cite {Barthelemy,Rozenfeld,Manna-Sen}.

       In this paper we studied the effect of a global directional 
    preference on the statistics of scale-free networks embedded in the Euclidean 
    space. A typical link in this model must have a positive component
    along some preferred direction. Similar to the
    directed versions of well known models of Statistical Physics \cite {DP,DSM,DRN,DSAW}
    our spatially directed networks have
    different universal critical behaviour compared to their undirected counterparts.

      A two-dimensional network is grown whose nodes are the points at randomly 
   selected positions within
   an unit square on the vertical $x-y$ plane. To construct a network of $N+1$
   nodes, let $(x_0, x_1, x_2, ... , x_N)$ and $(y_1, y_2, ... , y_N)$ be the $2N+1$
   independent random variables identically and uniformly distributed within 
   the interval $\{0,1\}$. Let a specific set of values of the random variables 
   $\{(x_0,0),(x_1,y_1),(x_2,y_2), ... , (x_N,y_N)\}$ represent the co-ordinates 
   of the of $N+1$ randomly distributed points. The growth of the network starts 
   with only one node $(x_0,0)$ on the bottom side of the unit square and then
   the other nodes are added one by one at unit rate according to their serial 
   numbers $i=1$ to $N$.

\begin{figure}[top]
\begin{center}
\includegraphics[width=8.5cm]{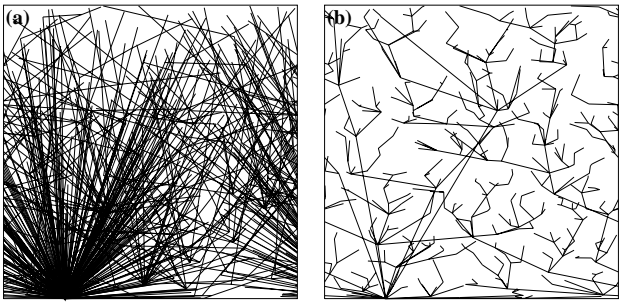}
\end{center}
\caption{
Pictures of the networks generated from the same distribution of 513 points
within the unit box. A large degree node is visible for a DSFN in (a) and 
long length early links are observed for a DMGN in (b). 
}
\end{figure}

      We assume that the global directional bias is the gravity and acts along 
   the $-y$ direction which restricts the choice of the link: a new node can only 
   be connected to a node positioned below this node. In practice 
   when the $t$-th node is introduced, we consider the subset $S_t$ of the nodes 
   situated below the $t$-th node. The $t$-th node is then connected to any node 
   of this subset using some specific attachment rule. In addition, we assume that 
   from each node only one link comes out but any number of links can terminate 
   on this node. This condition ensures that the network is a singly connected 
   tree graph. Initially the 0-th point is assigned the degree $k_0(0)=1$. Link
   lengths are measured using the periodic boundary condition imposed only along 
   the $x$ direction because of the anisotropy. Depending on how a node from the 
   sub-set $S_t$ is selected for connection we consider the following two models:

   (a) {\it Directed scale-free Network (DSFN):} The $t$-th node is randomly 
   connected to a node $i$ of the subset $S_t$ using an attachment probability 
   which is linearly proportional to the degree $k_i(t)$ of the node $i$ at time
   $t$ as: $\pi_i(t) \sim k_i(t)$.
  
   (b) {\it Directed minimal growing network (DMGN):} The $t$-th node is connected 
   with probability one to the nearest node in the subset $S_t$. Pictures of typical
   network configurations are shown in Fig. 1.

\begin{figure}[top]
\begin{center}
\includegraphics[width=6.5cm]{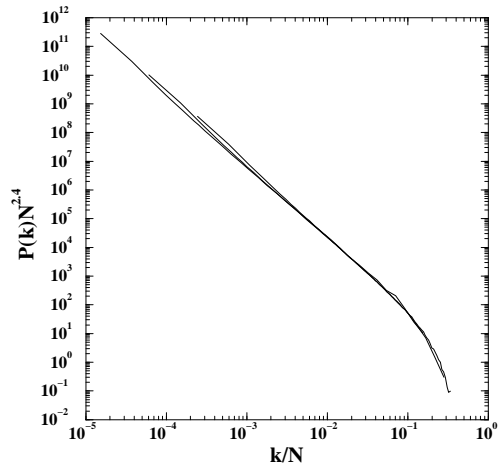}
\end{center}
\caption{
The scaled degree distribution for DSFN for network sizes
$N = 2^{12}, 2^{14}$ and $2^{16}$. The collapse of the data
at large $k$ values imply that the degree distribution 
exponent $\gamma \approx 2.4$.
}
\end{figure}

      A continuous tuning between these two different models is 
   possible by the choice of a suitable tunable parameter $\alpha$. This is achieved
   by modulating the attachment probability in DSFN by a link length $\ell$ 
   dependent factor like:
\begin {equation}
\pi_i(t) \sim k_i(t) \ell^{\alpha}.
\end {equation}
   This introduces a competition between the roles played by the degree as well as
   the link length on the attachment probability. The limiting extreme cases
   are the above two models. In the case with $\alpha$=0 the link lengths do not play
   any role and therefore the model corresponds to $DSFN$. On the other hand 
   when $\alpha=-\infty$ only the shortest link is selected with probability
   one irrespective of the degree of the node and therefore the model
   corresponds to $DMGN$. First we study these two limiting cases.

      For a scale-free network the nodal degree distribution has a power law tail:
   $P(k) \sim k^{-\gamma}$ and it obeys a finite size scaling form:
\begin {equation}
P(k,N) \sim N^{-\eta}{\cal F}(k/N^{\zeta}).
\end {equation}
   We numerically find that the degree distribution of DSFN (excluding the node on the
   bottom line) indeed follows such a scaling 
   form with $\eta \approx 2.4$ and $\zeta \approx 1$ (Fig. 2). This gives $\gamma_{DFSN}
   =\eta/\zeta \approx 2.4$. This value of $\gamma_{DSFN}$ is compared with $\gamma=3$ 
   for the BA model of SFN \cite {barabasi} and therefore it seems that DSFN
   belongs to a new universality class different from BA SFN. 
   On the other hand the degree distribution for the DMGN is found to decay exponentially
   as: $P(k) \sim exp(-\kappa k)$ with $\kappa \approx 0.74$. 

      For a tree graph, the branch size distribution is very important and the
   associated exponent may be used to characterize the graph. On a tree structure,
   each edge connects two branches of the tree. If an edge is selected randomly,
   the probability Prob$(s)$ that any one of the two branches supports $s$ nodes 
   also decays with a power law tail, Prob$(s) \sim s^{-\tau}$ and follows a 
   scaling form:
\begin {equation}
{\rm Prob}(s) \sim N^{-\eta_b}{\cal G}(s/N^{\zeta_b}).
\end {equation}
   For DSFN, we obtain $\eta_b \approx 2.15$ and $\zeta_b \approx 1$ which implies that
   $\tau_{DSFN} \approx 2.15$ compared to its exact value 2 for the BA scale-free
   network \cite {redner}. On the other hand for DMGN we find $\eta_b \approx 2$ 
   and $\zeta_b \approx 1$ so that $\tau_{DMGN} \approx 2$.

      The probability density distribution $D(\ell)$ gives the probability 
   $D(\ell) d\ell$ that an arbitrarily selected link has a length between
   $\ell$ and $\ell + d\ell$. For the undirected scale-free Euclidean 
   networks we saw that $D(\ell)$ has a power law variation $D(\ell) \sim 
   \ell^{\delta}$ \cite {yook1,Manna-Sen}. $D(\ell)$ can be calculated 
   exactly for both DSFN as well as DMGN in the following way. Let us try 
   to assign a link to the $(n+1)$-th node and denote $y=y_{n+1}$. Let $n_1$ 
   points be positioned below the $y$ level and $n_2=n-n_1$ points be above 
   this level. 

\begin{figure}[top]
\begin{center}
\includegraphics[width=6.5cm]{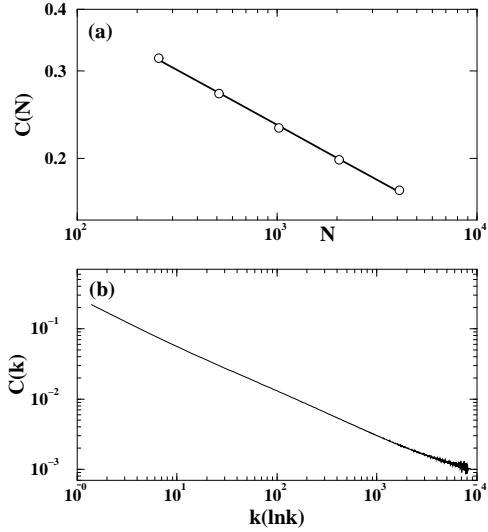}
\end{center}
\caption{
Variations of the average clustering co-efficients for DSFN: (a) Over the
whole network, $C(N) \sim N^{-\beta_N}$ and (b) Over the subset of
nodes having degree $k$ only $C(k) \sim [k(\ln k)]^{-\beta_k}$. Our estimates
are $\beta_N \approx 0.23$ and $\beta_k \approx 0.64$.
}
\end{figure}

      We first calculate $D_{DMGN}(\ell)$. The probability that out 
   of $n_1$ nodes the node which is nearest to the $(n+1)$-th node 
   is positioned at a distance between $\ell$ and $\ell+d\ell$ has 
   two distinct contributions: 
   One from the case of all $y > \ell$ and the other for all $y < \ell$ 
   (due to the presence of the boundary 
   at $y=0$). For the first case, the probability of the $(n+1)$-th point 
   being at a particular height $y$ is given by 
\begin{eqnarray}
  &   & \pi\ell d\ell \Sigma^{n}_{n_1=0} \{ ^{n}C_{n_1}\}n_1 [y-\pi\ell^2/2]^{n_1-1}(1-y)^{n_2} \nonumber \\
  & = & \pi\ell d\ell (n) [1-\pi \ell^2/2]^{n-1}. \nonumber
\end{eqnarray}
   Here the weight factor from each partitioning of $n$ into $n_1$ and 
   $n_2$ has been taken into account. The above probability after 
   integration over $y$ from $\ell$ to $1$ and summed over all 
   $n$ from 0 to $\infty$ gives the net contribution to $D(\ell)$ for all $y>\ell$:
\begin{eqnarray}
 D_{DMGN}(y>\ell)& = & \pi\ell [1-\ell] \Sigma^{\infty}_{n=0}n [1-\pi\ell^2/2]^{n-1} \nonumber \\
                 & = & \frac {4}{\pi\ell^3} (1-\ell). \nonumber
\end{eqnarray}
   Similarly the probability that the $(n+1)$-th point is at a specific height $y<\ell$ is
\begin{eqnarray}
 & & 2\ell sin^{-1} \frac{y}{\ell} d\ell \Sigma^{n}_{n_1=0}\{^{n}C_{n_1}\}(n_1) \nonumber \\
 & &\times [y-(\ell^2 sin^{-1}\frac{y}{\ell}+y\sqrt{\ell^2-y^2})]^{n_1-1}(1-y)^{n_2} \nonumber \\
 &=&2\ell sin^{-1} \frac{y}{\ell}n  [1-(\ell^2 sin^{-1} \frac{y}{\ell} + y\sqrt{\ell^2 -y^2})]^{n-1}d\ell. \nonumber 
\end{eqnarray}
   As before a similar sum over $n$ and integration over $y$ from $y = 0$ to $y=\ell$ 
   in the above expression gives the following contribution to $D(\ell)$ for all $y<\ell$:
\begin{eqnarray}
D_{DMGN}(y<\ell) & = & \frac{A}{\ell^2} \nonumber \\
{\rm where} \quad A & = & \int_0^{1} \frac{2sin^{-1}z}{{sin^{-1}z}+z\sqrt{1-z^2}}dz. \nonumber
\end{eqnarray}
   Hence the total distribution is given by:
\begin{equation}
   D_{DMGN}(\ell) = \frac {4}{\pi \ell^3} (1-(1-\frac{\pi}{4} A)\ell).
\end{equation}

\begin{figure}[top]
\begin{center}
\includegraphics[width=6.5cm]{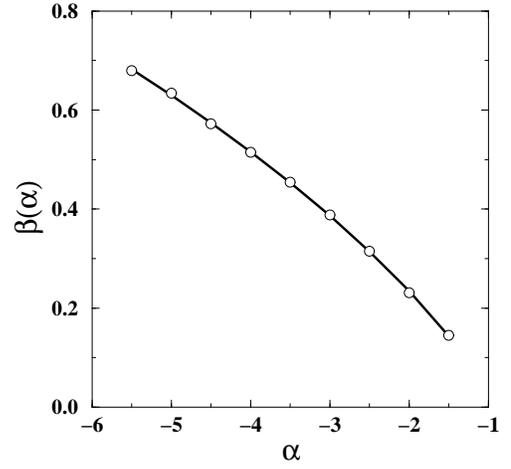}
\end{center}
\caption{
Variation of the exponent $\beta(\alpha)$ characterizing the
stretched exponential degree distribution. The continuous
curve is a fit to the data points (circles) obtained by simulation 
to a form: $\beta(\alpha) = a(-\alpha^{\nu})-b$ such that
$\beta$ is extrapolated to zero at $\alpha_c \approx -0.85$.
}
\end{figure}

      For the DSFN also one has:
\begin{eqnarray}
D_{DSFN}(y>\ell) & = & C\int_\ell ^1 \frac {\pi \ell}{y}dy \nonumber \\
{\rm and} \quad D_{DSFN}(y<\ell) & = & C \int_0 ^\ell \frac {2 sin^{-1}(y/\ell)}{y}dy \nonumber
\end {eqnarray}
   where $C$ is a constant.
The total distribution is therefore given by
\begin{eqnarray}
D_{DSFN}(\ell) & = & C(B \ell - \ell\ln (\ell)) \\
{\rm where} \quad B & = & 2\int_0^1 \frac {sin^{-1}z}{z} dz. \nonumber
\end{eqnarray}
From our numerical calculations we estimate $C \approx 1.59$ and $B \approx 1.86$.
For any non-zero $\alpha$, the corresponding distribution can be obtained
by simply multiplying the above expression by $\ell^\alpha$.

      The clustering co-efficient $C(N)$ of a network of $N$ nodes measures the 
   the local correlations among the links of the network. More precisely it 
   measures the probability that two neighbours of an arbitrary node are
   also neighbours. If the $i$-th node has the degree $k_i$ and there are 
   $e_i$ links among the $k_i$ neighbours of $i$ then the clustering 
   co-efficient of the site $i$ is: $C_i = 2e_i / [k_i(k_i-1)]$ whereas the 
   clustering co-efficient of the whole network is: $C(N) = \langle C_i 
   \rangle$. For a number of networks it has been observed that the clustering 
   co-efficient decreases with $N$ like $C(N) \sim N^{-\beta_N}$ as the 
   network size $N$ increases. Also one can define a clustering co-efficient 
   $C(k)$ averaged over the subset of nodes of degree $k$ on the network. It has 
   been also observed that $C(k) \sim k^{-\beta_k}$ for some networks. We 
   estimated these exponents for DSFN and found that $\beta_N \approx 0.23$ 
   whereas $C(k)$ has a logarithmic modulation like $C(k) \sim [k(ln k)]
   ^{-\beta_k}$ with $\beta_k \approx 0.64$ (Fig. 3).

      Finally we study the variation of the degree distribution $P(k)$ with
   the parameter $\alpha$. For finite negative values of $\alpha$
   the distribution fits very well to a stretched exponential form:
   $P(k) \sim \exp(-ck^{\beta(\alpha)})$ where $\beta(\alpha)$ is 
   expected to reach to one as $\alpha \to -\infty$ and to zero
   as $\alpha \to \alpha_c$. Our numerical estimates for $\beta$ have
   been plotted in Fig. 4 with $\alpha$ and this data fits very well
   to form $\beta(\alpha) = a(-\alpha)^{\nu}-b$ where the constants
   are estimated to be $a \approx 0.47$, $\nu \approx 0.51$ and $b \approx 0.43$. This
   implies that the stretched exponential form continues to be
   valid till $\alpha=\alpha_c$ where $\beta=0$ and beyond that
   the degree distribution is a power law. Though from the values of 
   $a, \nu$ and $b$, $\alpha_c$ is estimated to be $-0.85$ we believe
   $\alpha_c=-1$ is more plausible. Also our numerical results indicate
   that for all $\alpha > \alpha_c$ the degree distribution exponent 
   $\gamma$ maintains its value of $\alpha=0$.
   
      To summarize, we studied the directed version of the 
   Barabasi-Albert scale-free network grown on a two-dimensional vertical
   plane. Our numerical results on degree as well as branch size
   distributions indicate that this network
   belongs to a new universality class compared to its 
   undirected version. A competition between the degree $k$ of the 
   nodes and a link length dependent factor $\ell^{\alpha}$ in the 
   attachment probability is seen to control the network
   behaviour. In the limit $\alpha \to -\infty$ one gets the
   directed minimally growing network with exponentially decaying
   degree distribution. However for finite negative values of
   $\alpha$ stretched exponential distributions are observed.
   The link length distribution is calculated analytically for all
   values of $\alpha$.

   GM thankfully acknowledged facilities at S. N. Bose National Centre 
   for Basic Sciences. PS acknowledges financial support from DST, Grant No. 
   SP/S2-M11/99.

\leftline {Electronic Address: manna@boson.bose.res.in}

\end {document}